\documentclass[aps,prl,showpacs,floatfix,twocolumn]{revtex4}

\usepackage{amsmath}
\usepackage{amsfonts}
\usepackage{amssymb}
\usepackage{bm}
\usepackage{color}
\usepackage{graphicx}

\begin{document}

\title{Role of Committed Minorities in Times of Crisis}
\author{Malgorzata Turalska, Bruce J. West, Paolo Grigolini}
\affiliation{Center for Nonlinear Science, University of North Texas, P.O. Box 311427,
Denton, Texas, 76203, USA}
\date{\today}

\begin{abstract}
We use a Cooperative Decision Making (CDM) model to study the effect of
committed minorities on group behavior in time of crisis. The CDM model has
been shown to generate consensus through a phase-transition process that at
criticality establishes long-range correlations among the individuals within
a model society. In a condition of high consensus, the correlation function
vanishes, thereby making the network recover the ordinary locality
condition. However, this state is not permanent and times of crisis occur
when there is an ambiguity concerning a given social issue. The correlation
function within the cooperative system becomes similarly extended as it is
observed at criticality. This combination of independence (free will) and
long-range correlation makes it possible for very small but committed
minorities to produce substantial changes in social consensus.
\end{abstract}

\maketitle

Understanding the influence that committed minorities exert on both local
and global politics is an issue of overwhelming importance \cite{REFone}.
The surprising social phenomena of the Arab Spring \cite{clive} and the
Occupy Wall Street movement \cite{occupy} posit the question of whether the
active role of committed groups may produce political changes of significant
importance. Under what conditions are the convictions of a minority going to
dominate the future direction of a society? \textit{Couzin et al. \cite%
{couzin11} present experimental evidence that a committed minority of fish
can override the consensus of a substantially larger school of fish.}

Recent studies demonstrate that the abrupt, discontinuous events such as
catastrophic failures in power grids, computer networks and financial
markets crashes originate from the high connectivity of those networks \cite%
{REFfour}. Consecutively, since local interactions between building blocks
of such networks play an analogous role to the interactions in solid state
matter, adoption of phase-transition perspective to describe complex network
dynamics \cite{REFfive} comes as a natural extension of statistical physics
to the field of network science. As early as 1975 Haken \cite{haken} used
the concept of phase transition to interpret the 1968 French student
revolution. This approach explained the remarkably rapid transition from
traditional morality to sexual liberation, but without investigating the
role the minority of protesting students may have played in triggering this
change. Only recently the concept of inflexible agents $($committed
minorities$)$ who retain their opinion regardless of their social
environment has been introduced by Galam and Jacob \cite{galam}. Xie \emph{%
et al.} \cite{korniss} studied the influence of these inflexible agents and
found that when this committed minority reaches the threshold of $10\%$ the
opinion of the entire social network can be reversed.

The main purpose of this Letter is to demonstrate that the abrupt changes in
the organization of social groups, rather than being moments of disorder,
are instances of increased spatial correlation between the units of the
network. This condition of extended cooperation, similar to the critical
state of a phase transition, allows for a small subgroup of the society to
exert noticeable influence over the whole system. As a starting point we
consider the Cooperative Decision Making (CDM) model of two-state units $%
s_{i}$, each of which represents an agent making decision to agree ("yes")
or disagree ("no") on a given issue and whose dynamics are described by a
master equation on a two-dimensional lattice \cite{gosia1,gosia2}. In the
absence of interactions the probability for a given unit to change its
decision from "yes" to "no", or vice versa, is given by a Poisson
distribution with a transition rate $g<1$. When the interaction between
agents is turned on, the probability that a given unit is going to change
its decision becomes time dependent, yielding transition rates
\begin{equation}
p_{i\rightarrow j}(t)=g\exp [K\{M_{i}(t)-M_{j}(t)\}/M]  \label{prob1}
\end{equation}%
where $M_{i}$ is the number of nearest neighbors in the state $i=\{$"yes",
"no"\}, and $M$ is the total number of nearest neighbors. Global decisions
of a network composed of $N$ units whose interactions are described by Eq. %
\ref{prob1} can be defined by the time-dependent global order parameter $\xi
(t)=\left\vert N_{yes}(t)-N_{no}(t)\right\vert /N$, where $N_{yes}$ and $%
N_{no}$ are the global counts of units being in one of the two states at a
given time $t$. Here we consider the units of the CDM model to be on nodes
of a two-dimensional lattice with periodic boundary conditions. \textit{In
the simulation each element is updated in a given time step after which the
transition rates in Eq. \ref{prob1} are updated for the next time step.}

Similarly to the well known model proposed by Vicsek \cite{vicsek} to study
the cooperative dynamics of swarms of birds, the CDM model demonstrate the
shift from a configuration dominated by randomness to an organized state
once the cooperation parameter is increased above the critical value $K_{c}$%
. For values of the coupling parameter $K$ corresponding to the disorganized
phase $(K<K_{c})$, single units are only weakly influenced by the decisions
of their neighbors and change their state with probability only slightly
larger than the decoupled rate $g$. Thus, the fluctuations of the global
order parameter $\xi (t)$ are characterized by small amplitude and very fast
oscillations about the zero-axis [see Fig.1a]. For $K>K_{c}$, the
interaction between units gives rise to a majority state, during which a
significant number of agents adopt the same opinion at the same time [see
Fig.1c].

\begin{figure}[t]
\includegraphics[width=8.6 cm]{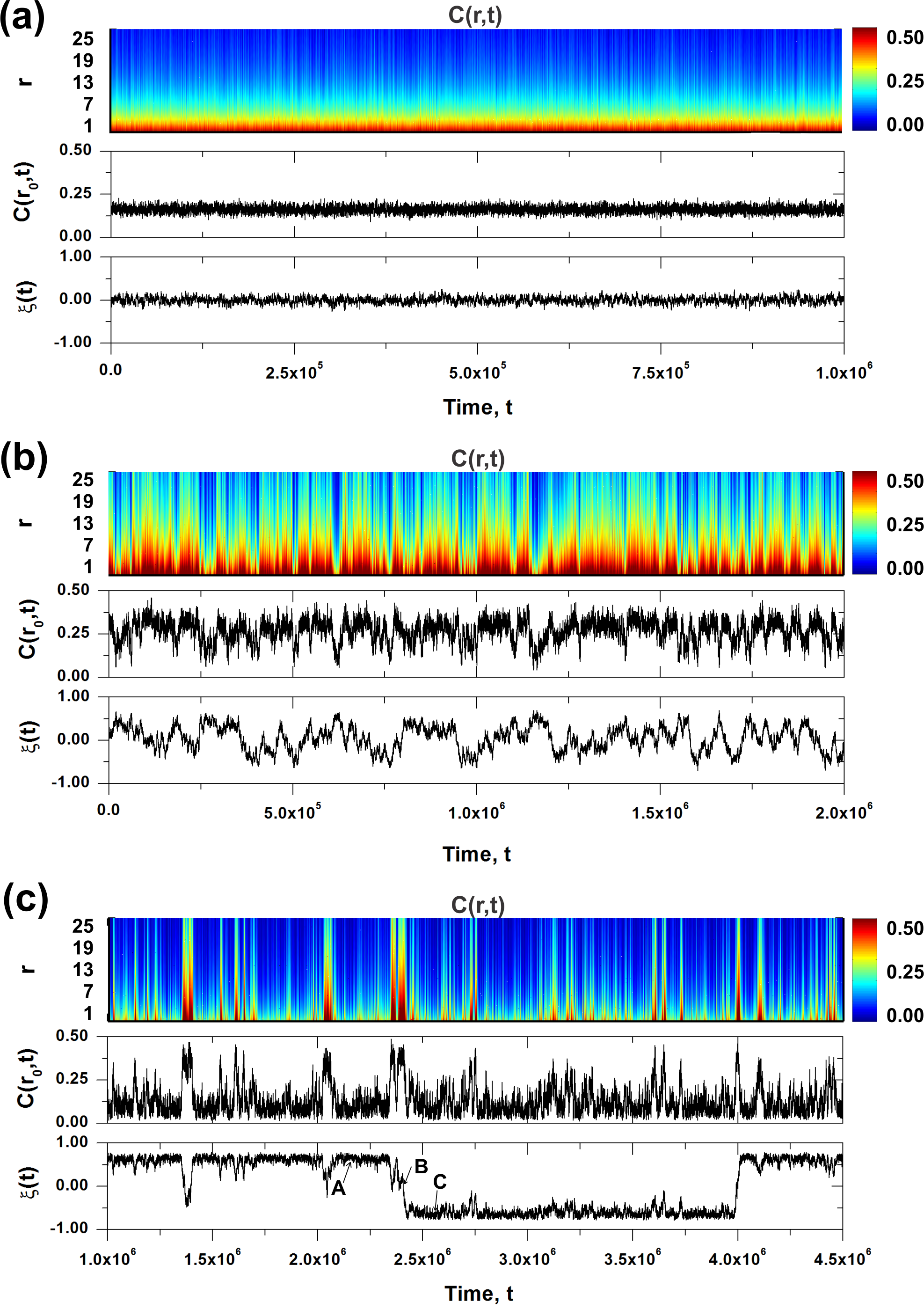}
\caption{Temporal evolution of the correlation function $C(r,t)$. On each
panel color map encodes spatiotemporal variability of correlation $C(r,t)$,
middle plot shows cross-section of this map for $r_{0}=15$ and bottom plot
shows corresponding evolution of the global order parameter $\protect\xi (t)$%
. Analysis performed for $K=1.50$ (a), $K=1.62$ (b) and $K=1.66$ (c).
Lattice size is $N=100$x$100$ nodes and transition rate $g=0.10$. }
\label{fig 1}
\end{figure}

We quantify the changes in temporal properties of the order parameter $\xi
(t)$, which accompany the phase transition, by evaluating the waiting-time
distribution density $\psi (\tau )$ and survival probability $\Psi (\tau
)=\int_{\tau }^{\infty }dt^{\prime }\psi (t^{\prime })$ of time intervals $%
\tau $ between consecutive recrossing of the zero-axis. As illustrated in
Fig. 2, in the subcritical regime $\Psi (\tau )$ has an exponential form,
which reflects the large independence of single units from their neighbors.
For $K>K_{c}$, the abrupt transitions between consecutive majority
intervals, that result from lattice having finite size, are responsible for
an exponential shoulder present in $\Psi (\tau )$. It is important to notice
that for a given value of the transition rate $g$, departing significantly
from the limiting condition $g\rightarrow 0$, there is no theoretical
prediction for the critical coupling strength $K_{c}$ \cite{gosia2}. The
value of $K_{c}$ estimated with traditionally used Binder cummulant method
\cite{REF13} yields $K_{c}^{B}=1.644$ when $g=0.10$. However, since this
approach estimates critical coupling for a network of infinite size, it is
not surprising that $\Psi (\tau )$ evaluated for $K_{c}^{B}$ and a lattice
of size $N=100\times 100$ nodes shows an exponential shoulder, which is a
hallmark of the organized phase. Simultaneously, following \cite{fabio} we
observe that for a network of finite size one can always find a value of the
interaction strength parameter $K$ for which the exponential shoulder
vanishes and for which $\Psi (\tau )$ is an inverse power-law function. We
assume this condition to correspond to the critical point for a network of
finite size, and we obtain $K_{c}=1.625$ for $N=100\times 100$.

\begin{figure}[t]
\includegraphics[width=8.6 cm]{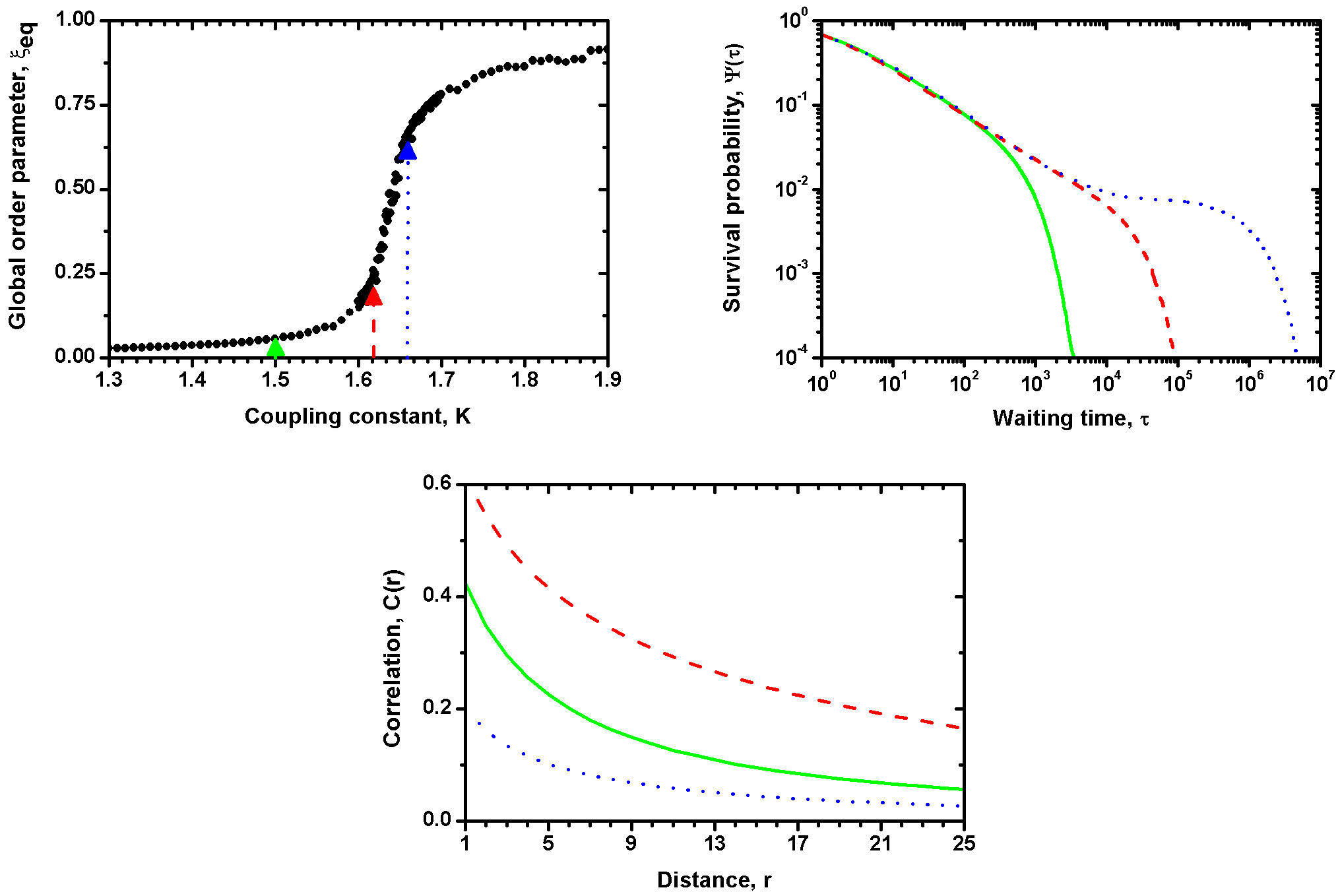}
\caption{Phase transition properties of the CDM model. (Top left panel)
Phase transition diagram for the amplitude of the global order parameter $%
\protect\xi (t)$ as a function of the coupling parameter $K$. (Top right
panel) Survival probability distribution $\Psi (\protect\tau )$ for selected
values of $K$: solid line corresponds to $K=1.50$, dashed line to $K=1.62$
and dotted line to $K=1.66$. (Bottom panel) Correlation $C(r)$ as a function
of the Euclidean distance $r$ between nodes of the lattice for $K$ being $%
1.50$, $1.62$ and $1.66$, respectively. Lattice size is $N=100$x$100$ nodes,
transition rate $g=0.10$. }
\label{fig 2}
\end{figure}

Further insight into the dynamic properties of the CDM model is obtained by
adopting the statistical measure of spatial influence, that being the
correlation function \cite{REF14} $C(r)$ between the nodes separated by the
Euclidean distance $r$:
\begin{equation}
C(r)=\langle s_{i}s_{i+r}\rangle -\langle s_{i}\rangle \langle
s_{i+r}\rangle .  \label{corr}
\end{equation}%
The quantity $\langle s_{i}s_{i+r}\rangle $ denotes an average over all
pairs of units on the lattice separated by distance $r$ and $\langle
s_{i}\rangle =\langle s_{i+r}\rangle $ denotes averages over all the units
of the network. Figure 2 shows $C(r)$ for selected values of the coupling
constant $K$. For both sub-critical ($K<K_{c})$ and super-critical ($K>K_{c})
$ values of the interaction strength, the correlation function $C(r)$
decreases rapidly as a function of the distance between nodes. However, at
criticality, $K=K_{c}$, we observe the correlation length to be
significantly more extended than either the sub- or super-critical cases, a
characteristic property of systems at a phase transition \cite{REF15}. It is
important to note that since we consider a network of finite size, this
extended correlation implies the emergence of dynamical coupling between
units that are not nearest neighbors and therefore not directly linked.

\begin{figure}[t]
\includegraphics[scale=0.95]{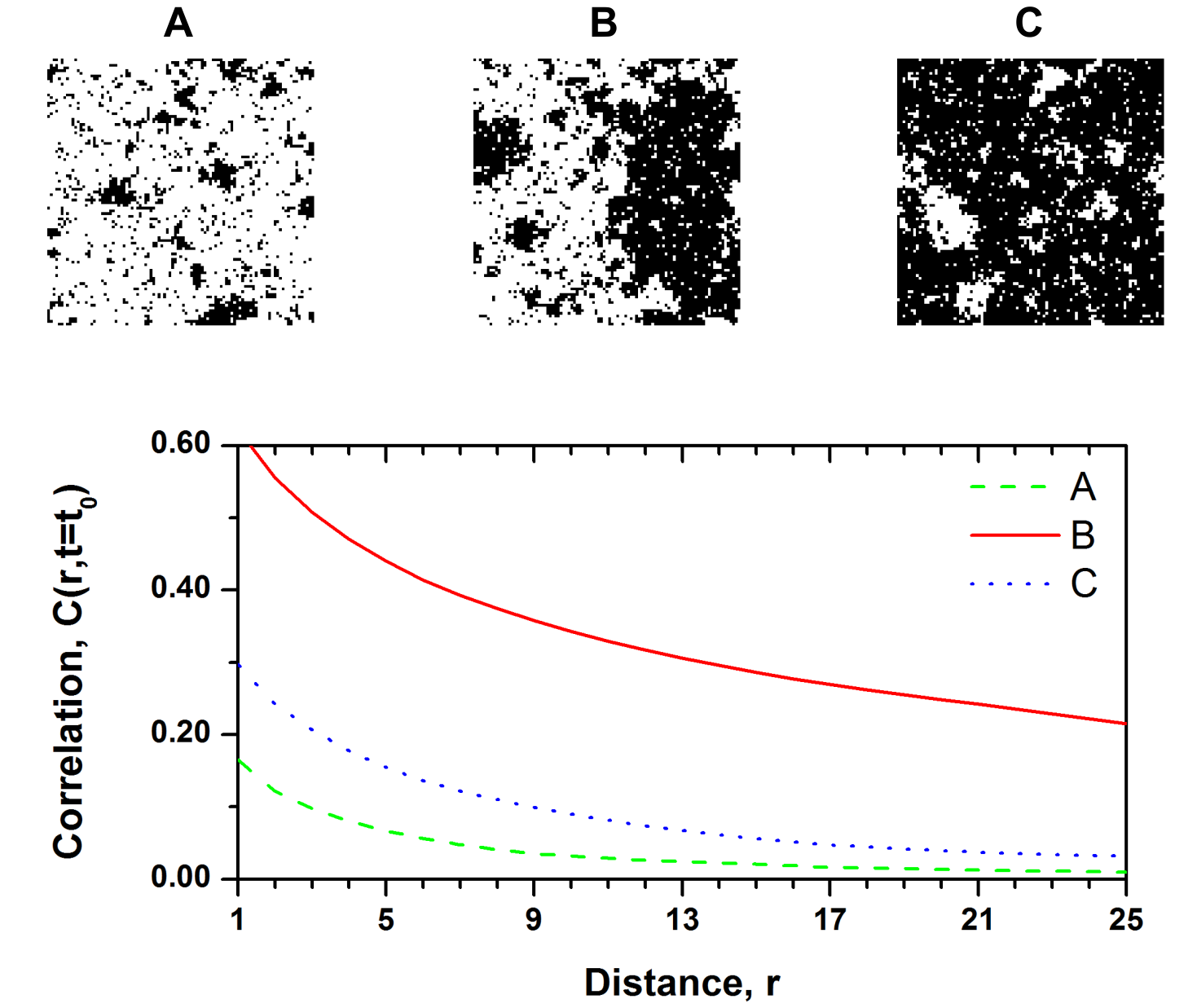}
\caption{ Configuration of the lattice (a-c) and corresponding correlation
function (bottom panel) for instances marked on bottom panel of Fig. 2c by
arrows. White areas correspond to the units in state "yes" and black to the
units in the state "no". Lattice size is $N=100\times 100$, coupling
constant is $K=1.66$ and transition rate is $g=0.10$.}
\label{fig3}
\end{figure}

Although traditionally the correlation function $C(r)$ is considered an
equilibrium property of a network \cite{REF15}, the central result of this
Letter is obtained by modifying Eq. \ref{corr} in order to study temporal
fluctuations of the correlation function. For each instant of time we define
the spatiotemporal correlation function
\begin{equation}
C(r,t)=\langle s_{i}(t)s_{i+r}(t)\rangle -\langle s_{i}(t)\rangle \langle
s_{i+r}(t)\rangle   \label{corrT}
\end{equation}%
and as shown on Fig. 1, we observe that the fluctuations of the correlation
function $C(r,t)$ closely track those of the global order parameter $\xi (t)$%
. In particular, the abrupt transitions between epochs of dominating
majority, the instances of crisis that take place when $K>K_{c}$, correspond
to jumps in $C(r,t)$. Additionally, as demonstrated in Fig. 3, careful
inspection of the network at the instant of crisis reveals an extended
correlation length when compared with the organization of the lattice during
majority rule. This observation provides an explanation of how committed
minorities succeed in their goal of inducing significant social change.
Since at the moment of the jump the order parameter vanishes, $\xi (t)=0$,
we interpret those events as \emph{free will} states. One could expect that
the network had lost its organization and is randomly configured at that
moment. However, such an interpretation would signify the presence of only
local coupling and therefore could not explain how a small bias exerted on
the network is able to produce large-scale changes, as shown subsequently.

\begin{figure}[t]
\includegraphics[scale=0.95]{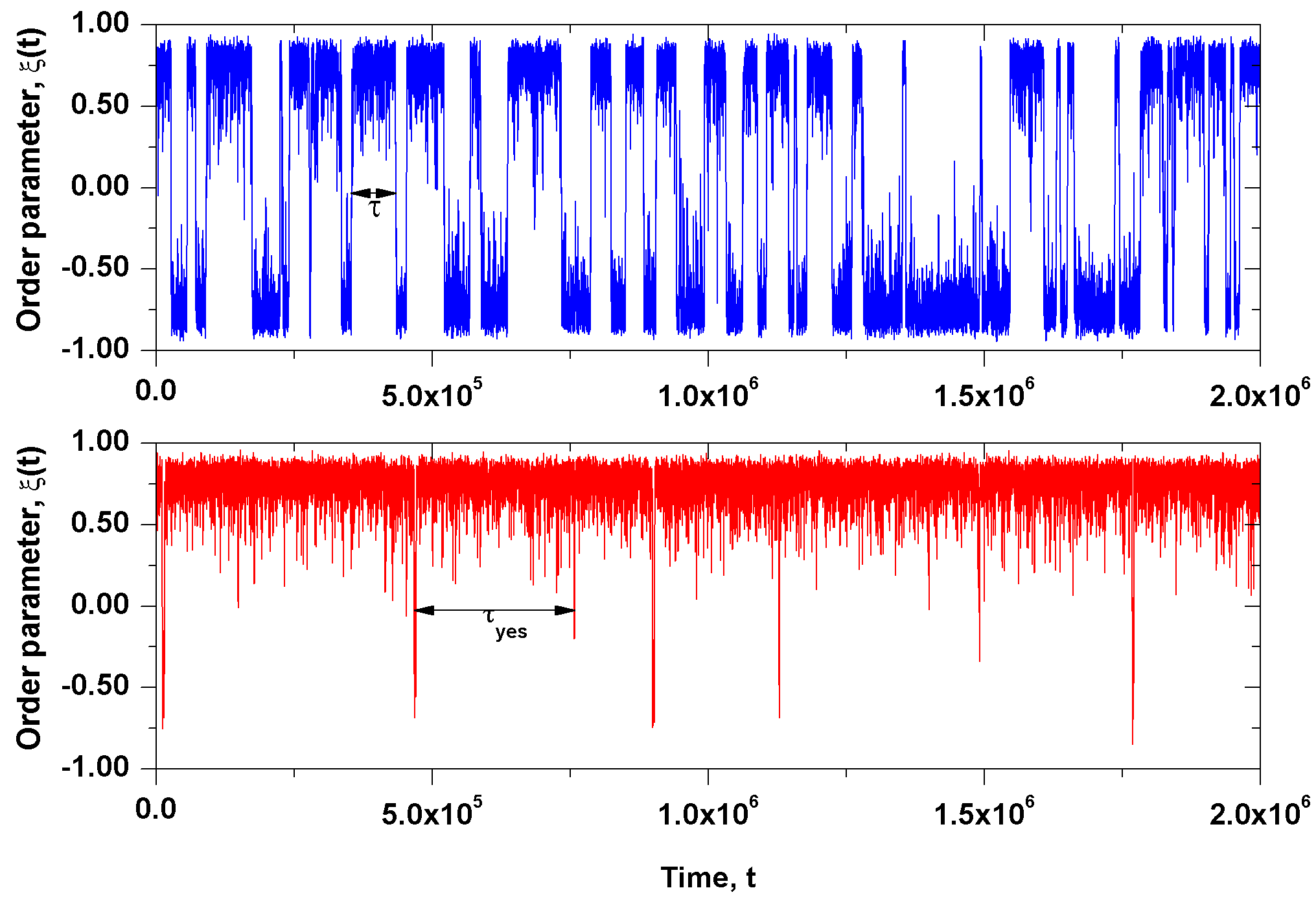}
\caption{ Small number of nodes maintaining constant opinion influence
significantly the behavior of the system in the organized phase. (Top panel)
Fluctuations of the global order parameter $\protect\xi (t)$ for $K=1.65$
and lattice of size $N=100\times 100$ nodes. (Bottom panel) The behavior of $%
\protect\xi (t)$ once $1\%$ of the randomly selected nodes are kept in state
"yes" at all time. Transition rate is $g=0.10$ for both.}
\label{fig4}
\end{figure}

Given the influence of today's social events, that more and more frequently
affect global political and economical scene, now is the right time to study
the impact of committed minorities on a society. A member of the committed
minority considered herein is a randomly selected element on the lattice
that keeps its decision of either "yes" or "no" independently of the opinion
of its neighbors. To prove that the committed minorities may operate
efficiently in spite of their very small number, in Fig. 4 we compare the
evolution of $\xi (t)$ in the absence of a committed minority to the
evolution of $\xi (t)$ in the presence of a relatively small $(1\%)$
committed group. In the case considered here a high value of the coupling, $%
(K>K_{c})$, leads to the extended condition of global consensus, during
which the influence of the minority is negligible. The rapidly decreasing
correlation function $C(r)$ reflects the rigidity of the network and
prevents the global transmission of the perturbation. However, from time to
time a crisis occurs where $\xi \left( t\right) =0$. In crisis the network
may undergo an abrupt change of opinion and the correlation length is
sufficiently large to make it possible for the committed minority to force
the social network to adopt their view. As a consequence, during the time
interval over which the minority acts it imposes its opinion over the whole
network.

\begin{figure}[t]
\includegraphics[scale=0.90]{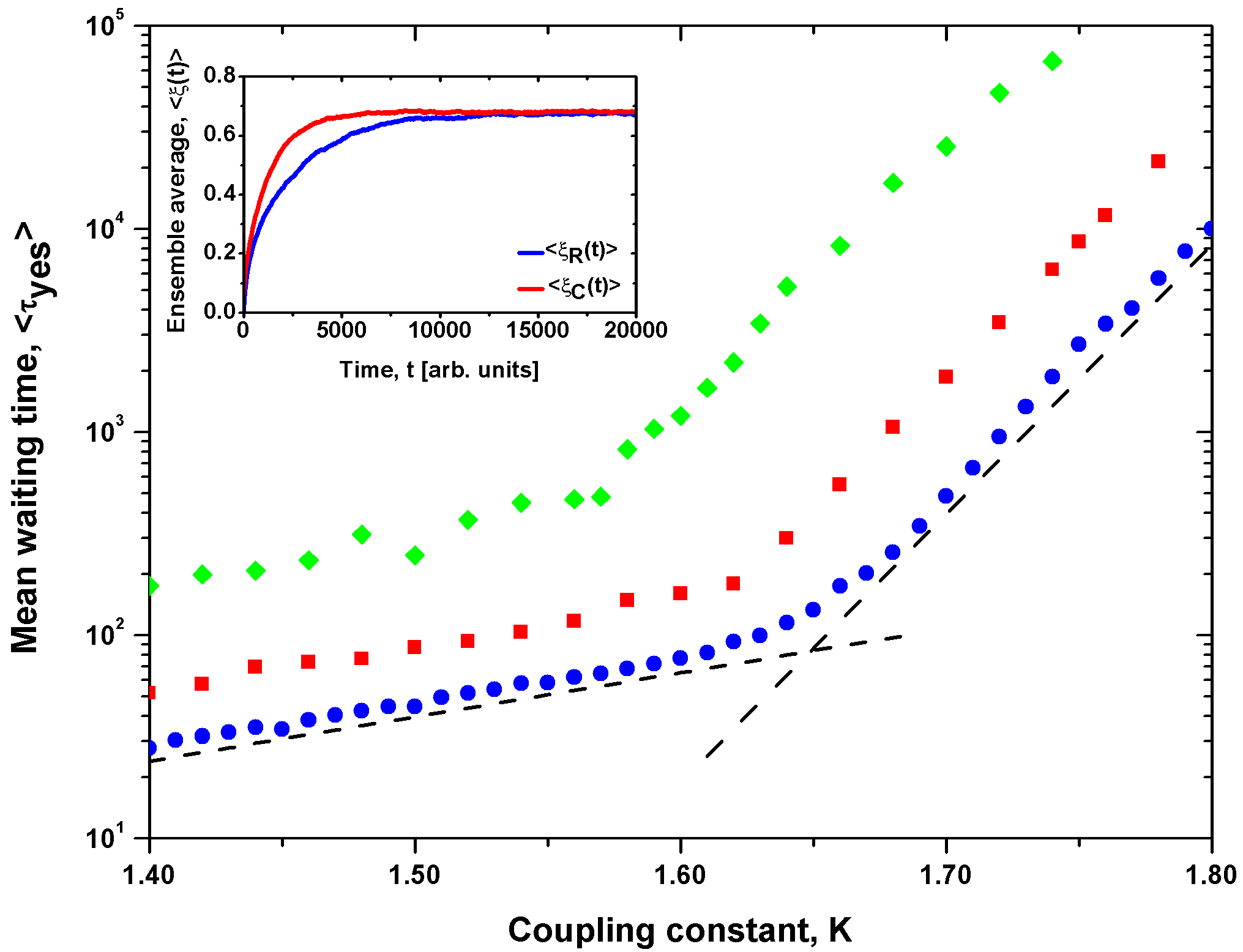}
\caption{ Mean waiting time $<\protect\tau _{yes}>$ obtained for system with
no acting minorities (blue dots) increases once $1\%$ (red squares) and $5\%$ (green diamonds) committed minority is present. Insert: The ensemble average of the CDM model realizations initialized with a random configuration (blue line) are compared with those initialized with a crisis configuration (red line). Lattice size is $N=20$x$20$, $K=1.65$ and $g_0=0.10$.}
\label{fig5}
\end{figure}

Finally, to quantify this phenomenon, we study how introducing a committed
minority affects the average lifetime of the consensus state, defined as $%
<\tau >=\int_{0}^{\infty }\Psi (\tau )d\tau $. It is important to note that
in the condition where no committed group is present the distribution of
time durations of global decision in "yes", $\Psi (\tau _{yes})$, and in
"no", $\Psi (\tau _{no})$, coincide and are equal to the distribution
evaluated for both time intervals, $\Psi (\tau )=\Psi (\tau _{yes})=\Psi
(\tau _{no})$. This symmetry is however broken once a minority is
introduced, as shown on Fig. 4, by an increase in temporal span of consensus
states that correspond to the opinion of the minority. In Fig. 5 we compare $%
<\tau >$ in the absence of the inflexible agents to $<\tau >$ in the
presence of a committed minority of sizes $1\%$ and $5\%$. First, in the
native case the average consensus time $<\tau >$ increases exponentially
with an increase in the coupling strength $K$, showing a faster rise once $%
K>K_{c}$, with a discontinuous change at the critical interaction strength.
This switch in the rate of increase confirms the validity of the approach
used to determine $K_{c}$ based on the temporal properties of $\xi (t)$.
Consecutively, the introduction of a small minority leads to a linear
increase in $<\tau >$ and the fact that two exponential regimes are
preserved confirms the crucial role that instances of crisis play in the
global transmission of minority opinion. To provide additional evidence to
support this claim, we consider an ensemble of realizations of the CDM
process: one initialized with random configuration and other with a
configuration derived from the instance of a crisis. Due to the presence of
a small minority each evolution of $\xi (t)$ spends more time in the state
"yes" and gives rise to a non-zero ensemble average $<\xi (t)>$. As shown on
the insert of Fig. 5, the long range correlations present at the time of
crisis cause the system to organize faster than happens when correlations
are strictly local.

Our approach does not allow us to confirm the observation made by Korniss et
al. \cite{korniss} that the minimal size of the committed minority necessary
to significantly affect the opinion of the entire network is 10\%. If we
assume that a substantial effect of a committed minority is defined by an
order of magnitude increase in the average consensus time $<\tau >$, Fig. 5
indicates that for CDM this requirement is realized by a committed minority
of less than $5\%$. Simultaneously, we observe that the effect of the
minority on the dynamics of the CDM model is preserved once the dimension of
the lattice is increased from two to three, which is contrary to the results
reported by Mobilia \cite{Mobilia}. We are convinced that those differences
are a consequence of the non-local interactions present in the CDM model and
the local, diffusion-like behavior shown by the models investigated in \cite%
{korniss, Mobilia}.

This Letter demonstrates that in a dynamical network the extended
correlation length emerges not only at criticality, but quite surprisingly
also at times of crisis which separate intervals of strong opinion. Our
results indicate that this property is responsible for the significant
efficiency of inflexible minorities in influencing entire networks. This
discovery may shed light onto the animal behavior recently observed \cite%
{couzin} in flocks of locust and reproduced with the cooperative model of
Vicsek \cite{vicsek}. In fact, the locust's change of direction seems to
occur in the supercritical condition and is accompanied by increased
activity of the single units. Simultaneously, it follows that there is no
need for specific organization of the members of the committed minority,
since due to increased correlation their impact is perceived even if the
inflexible units are randomly arranged within the network and are not in
direct contact with each other.

MT and PG gratefully acknowledge ARO for the financial support of this
research through grant W911NF-11-1-0478.


\begin{thebibliography}{99}
\bibitem{REFone} P. Ball, Nature, 480, 447-448 (2011).

\bibitem{clive} C. Barnett, Geoforum, 42, 263-265 (2011);  M. Al-Momani,
Middle East Law and Governance, 3, 159-170 (2011).

\bibitem{occupy} S. van Gelder (editor), \textit{This Changes Everything},
Berrett-Koehler Publishers, San Francisco (2011); N. Schneider, Nation, 293,
13 (2011); B. Moyer, Nation, 21, 11 (2011).

\bibitem{couzin11} I.D. Couzin et al., Science \textbf{334}, 1578-15-80
(2011).

\bibitem{REFfour} A. Vespignani, Nature, 464, 984-985 (2010).

\bibitem{REFfive} A. Barrat, M. Barthelemy, A. Vespignani, \textit{Dynamical
Processes on Complex Networks}, Cambridge University Press, 2008.

\bibitem{haken} H. Haken, Rev. Mod. Phys. 47, 67-121 (1975).

\bibitem{galam} S. Galam, F. Jacobs, Physica A, 366-376 (2007).

\bibitem{korniss} J. Xie et al., Phys. Rev. \textbf{84}, 011130 (2011).

\bibitem{gosia1} M. Turalska, M. Lukovic, B. J. West, P. Grigolini, Phys.
Rev. E \textbf{80}, 021110, 1-6, (2009).

\bibitem{gosia2} M. Turalska, B. J. West, P. Grigolini, Phys. Rev. E \textbf{%
83}, 061142, 1-6, (2011).

\bibitem{vicsek} T. Vicsek et al., Phys. Rev. Lett. \textbf{75}, 1226, 1-4,
(1995).

\bibitem{REF13} K. Binder, Z. Phys. B, 43, 119-140, (1981).

\bibitem{fabio} F. Vanni, M. Lukovi\'{c}, P. Grigolini, Phys.Rev.Lett.
\textbf{107}, 1-4, 078103 (2011).

\bibitem{REF14} T.T. Wu et al., Phys. Rev. B, 13, 316 (1976).

\bibitem{REF15} M. E. Fisher, Rep. Prog. Phys, 30, 615 (1967).

\bibitem{Mobilia} M. Mobilia, Phys. Rev. Lett. \textbf{91}, 028701, (2003).

\bibitem{couzin} C. A. Yates et al., PNAS, \textbf{106}, 5464-5469 (2009).
\end{thebibliography}
\end{document}